\documentclass[conference]{IEEEtran}

\makeatletter
\def\ps@headings{%
\def\@oddhead{\mbox{}\scriptsize\rightmark \hfil \thepage}%
\def\@evenhead{\scriptsize\thepage \hfil \leftmark\mbox{}}%
\def\@oddfoot{}%
\def\@evenfoot{}}
\makeatother
\newcommand{\resetcounters}{\setcounter{equation}{0}}
\usepackage{amsfonts,amsmath,amstext,verbatim}
\usepackage{amssymb}%
\usepackage{setspace}
\usepackage{ifpdf}

\makeatother  \renewenvironment{abstract}{%
  \small\bfseries\textit{Abstract}:  }

\ifCLASSINFOpdf
\usepackage[pdftex]{graphicx}
\graphicspath{{./figures/}{./}}
\usepackage[pdftex,colorlinks,
           bookmarks=true,%
           pdftitle=Stochastic\ Analysis\ of\ Non-slotted\ Aloha\ in\ Wireless\ Ad-Hoc\ Networks,%
           pdfauthor=B.\ Blaszczyszyn%
           \ -\  P.\ Muhlethaler,]{hyperref} 
\usepackage[numbers,sort&compress]{natbib}
\usepackage{hypernat}
\DeclareGraphicsExtensions{.pdf}
\else

\usepackage[dvips]{graphicx}
\graphicspath{{./figures/}{./}}
\DeclareGraphicsExtensions{.eps}
\usepackage[numbers,sort&compress]{natbib}
\fi

\newcommand{\md}{\text{\rm d}}

\newcommand{\E}{{\mathbf E}}
\newcommand{\Pro}{{\mathbf P}}
\newcommand{\ind}{1\hspace{-0.30em}{\mbox{I}}}

\renewcommand{\theequation}{\arabic{section}.\arabic{equation}}

\newtheorem{Th}{Theorem}[section]
\newtheorem{prop}[Th]{Proposition}
\newenvironment{Prop}{\bf\begin{prop}\rm\em}{\end{prop}} 
\newtheorem{res}[Th]{Result}
\newenvironment{Res}{\bf\begin{res}\rm\em}{\end{res}} 
\newtheorem{fact}[Th]{Fact}
\newenvironment{Fact}{\bf\begin{fact}\rm\em}{\end{fact}} 

\newcommand{\rem}{{\medskip\noindent \bf Remark:~\nolinebreak}}

\newcommand{\bfF}{\mathbf{F}}

\newcommand{\calL}{{\mathcal{L}}}
\newcommand{\calF}{{\mathcal{F}}}

\newcommand{\bbZ}{\Bbb{Z}}
\newcommand{\bbR}{\Bbb{R}}

\makeatother  \renewenvironment{abstract}{%
  \small\bfseries\textit{Abstract}:  }

\begin{document}
\title{Stochastic Analysis of Non-slotted Aloha in Wireless Ad-Hoc Networks}

\author{
\begin{tabular}{c c c}
Bart{\l}omiej B{\l}aszczyszyn & Paul M{\"u}hlethaler\\
INRIA/ENS and Math. Inst. Univ. of Wroc{\l}aw
&  INRIA  Rocquencourt\\
Paris FRANCE &  Le Chesnay FRANCE\\
Bartek.Blaszczyszyn@ens.fr & Paul.Muhlethaler@inria.fr
\end{tabular}
\vspace{-3ex}
}

\maketitle
\thispagestyle{empty} \pagestyle{empty}

\begin{abstract}
In this paper we propose two analytically tractable stochastic models of
  non-slotted Aloha for Mobile Ad-hoc NETworks (MANETs): one model
 assumes a static pattern of nodes while the other assumes that the
 pattern of nodes  varies over time.
Both models feature transmitters
randomly located in the Euclidean plane, according to a Poisson point
process with the receivers randomly located at a fixed distance from
the emitters.  We concentrate on the so-called outage scenario, where a
successful transmission requires a Signal-to-Interference-and-Noise Ratio
(SINR) larger than a given threshold.
With Rayleigh fading and the SINR averaged over the duration of the 
packet transmission, both  models lead to closed
form expressions for the  probability of successful transmission.
We show an excellent matching of these results with
simulations. Using our  models we  compare the performances of 
non-slotted Aloha to slotted Aloha studied in~\cite{JSAC}. 
We observe that when the path loss is not very strong
both models,  when appropriately optimized, exhibit similar
performance. For stronger path loss non-slotted Aloha performs worse
than slotted Aloha, however when the path loss exponent is equal
to~4 its density of successfully received packets
is still 75\% of that in the slotted scheme.
This is still much more than the 50\% predicted by the well-known 
analysis where simultaneous transmissions are never
successful. 
Moreover, in any path loss scenario, both schemes exhibit the same 
energy efficiency.
\end{abstract}


\section{Introduction}
Aloha is one of the most common examples of a multiple communication 
protocol; it is presented in many widely used books 
on data networks such as \cite{Schwartz87,Bertsekas01}.
A main characteristic of Aloha is its great simplicity: the core
concept consists in allowing each source to transmit a packet
and back-off for some random time before the next transmission
independently of other sources. This, of course, leads to collisions
and some packets have to be retransmitted. In order to evaluate the
fraction of packets that are transmitted successfully, a   
simple and widely used model assumes that simultaneous
transmissions are never successful. When the aggregate packet
transmission process  follows a Poisson distribution, the analysis of
this pure (or {\em non-slotted}) Aloha model   
shows that on average the fraction $1/(2e)\approx18.4\%$  of
successful transmissions can be attained, when the scheme is optimized  
(tuning the  mean back-off time). 
It also shows that this performance can be multiplied
by~2 in {\em slotted-Aloha}, when all the nodes are synchronized and
can send packets only at the beginnings of some universal time slots. 
This analysis,  which is very often taught to well exemplify 
the protocol's performance, is however based on the simple collision
model where two simultaneous transmissions always lead to a collision. 
This assumption is well adapted to wired networks but is not adequate
for wireless networks where {\em spatial reuse} is generally present.   

In this paper we analyze Aloha in a wireless 
network model featuring transmitters
randomly located in the Euclidean plane, according to a Poisson point
process, with the receivers randomly located at a fixed distance from
the emitters. We assume the SINR coverage context, which is
that where each successful transmission requires that the receiver be
covered by the transmitter with a minimum SINR.
We adopt a path loss model with a power-law mean signal-power decay
$l(u)=u^\beta$  on the distance $u$ and we assume some 
random independent fading model.
Whereas the analysis of slotted Aloha in such a MANET model 
has been done in~\cite{allerton}, in this paper we propose  
an analysis of non-slotted Aloha. 

As the {\em main theoretical contribution} of this paper
we build {\em two analytically tractable stochastic models of
  non-slotted Aloha}. We believe one of these models to be 
very close to the reality 
of a relatively static MANET, while the other 
assumes that the pattern of nodes 
is different at each transmission. With Rayleigh fading,
and when the interference is averaged in the  SINR packet reception constraint, 
both models lead to closed
form expressions for the {\em probability of   successful receptions}. 
The formula derived in the static MANET  requires numerical processing 
while the formula derived in the other model is more explicit.
Moreover,  we find an excellent matching of the values 
obtained for these two  analytical models.

We  also conduct extensive simulations of non-slotted
Aloha, both with averaged and maximal  interference
in the SINR constraint.
We show an excellent matching of both  analytical models with
the simulations in  the mean interference constraint case.   

Using our models we also compare the performances of 
slotted and non-slotted Aloha.
The {\em main findings of this analysis} show that:
\begin{itemize}
\item When the path loss exponent $\beta$ is small (close to its lower
  theoretical bound $\beta=2$) slotted and non-slotted Aloha (non-slotted 
with the mean interference constraint),
  when appropriately optimized, offer a similar space-time density of
  successful transmissions. 
\item For larger values of $\beta$, the optimized  
non-slotted Aloha gives a smaller  
density of  successful transmissions than the optimized slotted
Aloha, with the ratio asymptotically going to $0.5$ 
($\beta\to\infty)$
--- the value predicted 
by the widely used simplified model. However, for $\beta=4$ this ratio
is still~$75\%$.
\item When optimized, both slotted and non-slotted models exhibit the same energy
  efficiency (the mean number of successful transmissions per unit of energy
  spent) if one ignores the energy spent for maintaining 
  synchronization in the slotted scheme.
\end{itemize}

The rest of this paper is organized as follows. 
In the remaining part the current section
we recall some previous studies of Aloha.
Section~\ref{s.NetAloha} introduces 
the network model and our two models for non-slotted Aloha.  
Section~\ref{s.ST} contains the main stochastic analysis results of
this paper.
In Section~\ref{s.tuning} we show how the performance of the 
non-slotted Aloha can be optimized and compare it to the optimized
slotted Aloha.
Our conclusions are presented in Section~\ref{s.conclusion}.
In the Appendix  we present proofs of our mathematical results.

\subsection{Related Work} 
\label{ss.Relatedw}
Aloha and Time Division Multiple Access (TDMA) are the oldest multiple 
access protocol. Aloha, which is the ``mother'' of 
random protocols, was born in the early seventies, the seminal 
work describing Aloha~\cite{Aloha} being published in 1970. Aloha is 
very simple and also extremely widespread. 
Another very amazing characteristic of 
Aloha is that it has an extremely simple analysis which is widely
taught (cf. e.g \cite[4.2]{Bertsekas01}). 
However this analysis is only valid in the rough model where two 
simultaneous transmissions necessarily lead to a collision. 
Surprisingly, although Aloha was primarily designed to 
manage a wireless network, the first models for Aloha 
were more adapted to wired networks. To the authors' best knowledge 
the first contribution in which Aloha was explicitly studied in a wireless 
context is the paper by Nelson and Kleinrock~\cite{Nekl83}. 
The propagation model of this paper is very simple and it was only  
in 1988 that the widely referenced paper~\cite{verdu} was published, in 
which the first reasonable Aloha model for  wireless network with spatial 
reuse was proposed. 
However, even in that paper the spatial analysis of the protocol
remained simplified. More recently, in~\cite{BFS07}
a more refined spatial Aloha model was studied with local interactions
and the simplified collision model (where simultaneous transmissions 
always lead to collisions).
A little earlier in~\cite{allerton}, one
evaluated the probability of capture for Aloha in the SINR
outage context with Rayleigh fading, showing a direct link 
between the probability of capture and the Laplace transforms 
of the thermal noise and of the interference (also called shot-noise). 
The key factor of this analysis is the explicit formula of the 
Laplace transform of the interference created by a  Poisson pattern of
nodes.
This analysis of slotted Aloha  was completed in~\cite{JSAC}. 

The main contribution of the present  paper is the extension of the
analysis  of~\cite{allerton,JSAC} to pure (non-slotted) Aloha
allowing for a fair comparison of both schemes in the wireless MANET context.

There are many publications on the stability of 
Aloha and, more generally, random back-off protocols.
This problem is not addressed in the present paper.

\section{Network and Aloha Models}
\label{s.NetAloha}
\resetcounters
In this section we present comparable models of slotted and
non-slotted Aloha for wireless ad-hoc
networks. More precisely we first introduce the geometric 
model of the network. We then present the access schemes: 
slotted and non-slotted Aloha and the mathematical models 
used to represent these schemes. We also model 
the fading process and the external noise.  
For non-slotted Aloha, we propose a second model whose 
analysis is simpler. At the end of this section 
we show how to relate different model parameters to 
make a fair comparison of their performance.

\subsection{Location of Nodes --- The Spatial Poisson Bipolar Network
  Model} 
\label{ss.Bipolar}
We consider a {\em Poisson bipolar network model}
in which each point of the Poisson pattern 
represents a node of a Mobile Ad hoc NETwork (MANET) and is hence a potential transmitter.
Each node 
has  an associated receiver located at distance $r$. This receiver is 
not part of the Poisson pattern of points.

More precisely, using the formalism of the theory of point
processes, we will say that a snapshot of the MANET 
can be represented by an independently marked Poisson point process (P.p.p)
$\widetilde\Phi=\{(X_i,y_i)\}$,
where the {\em locations of nodes}  $\Phi=\{X_i\}$ form  a homogeneous P.p.p. on
the plane, with an intensity of $\lambda$ nodes per unit of space,
and where the mark $y_i$
denotes the location of the receiver for node $X_i$.
We assume here that no two transmitters have the same receiver
and that, given $\Phi$,  the vectors $\{X_i-y_i\}$ are i.i.d  with
$|X_i-y_i|=r$.~\footnote{%
The fact that all receivers are at the same distance from their
transmitters is a simplification. There is no
difficulty extending what is described below to the case where these
distances are independent 
and identically distributed random variables, independent of
everything else. A further possible extension assumes that  
the actual receivers are selected from some, say Poisson, process
of potential receivers, common for all MANET nodes, taking the nearest 
point to the emitter. A more involved model assumes that the 
transmitters of the MANET  choose their receivers in
the original set~$\Phi$ of nodes of the MANET;
see~\cite[Chapter~17]{FnT2}
for precise descriptions and the analysis of slotted Aloha.}  

\subsection{Aloha Models --- Time  Added} 

We will now consider two time-space scenarios appropriate for slotted
and non-slotted Aloha. 
In both of them the planar locations of MANET
nodes and their receivers $\widetilde\Phi$ {\em remain fixed}.
It is the medium access (MAC) 
status of these nodes that will evolve differently 
over time depending on which of the following two models is used.

\subsubsection{Slotted Aloha}
\label{sss.slotted}
In this model we assume that the time is discrete, i.e. divided into
slots of length $B$ (the analysis will not depend on the length of
the time-slot) and labeled  by integers $n\in\bbZ$. 
The nodes of $\Phi$ are {\em perfectly synchronized} 
to these (universal) time slots  and send packets according to the
following {\em slotted Aloha:
each node, at each time slot independently tosses a coin with 
some bias $p$ which will be referred to as the medium access
probability (MAP); it sends the packet in this time slot if the outcome is
heads and  backs off its transmission otherwise}.
This evolution of the MAC status of each node $X_i$ can be formalized
by introducing its further (multi-dimensional) mark $(e_i(n)
:n\in\bbZ)$,
where  $e_i(n)$  is the medium access indicator of node $i$ at time
$n$; $e_i(n)=1$ if node $i$ is allowed to transmit in the time
slot considered and 0 otherwise.
Following the Aloha principle we assume that $e_i(n)$ are hence
i.i.d. (in $n$ and $i$) and independent
of everything else, with $\Pro(e_i(n)=1)=p$.
We treat $p$ as the main parameter to be tuned for slotted Aloha.
We will call the above case {\em the slotted Aloha model}.

\subsubsection{Poisson-renewal Model of Non-slotted Aloha}
\label{sss.renewal}
In this {\em non-slotted Aloha} model all the nodes of $\Phi$
independently, without synchronization, 
send packets of the same duration $B$ 
and then back off for some random time.
This can be integrated in our model by introducing marks
$(T_i(n):n\in\bbZ)$,
where  $T_i(n)$ denotes  the beginning of the $n\,$th transmission
of node $X_i$ with
$T_i(n+1)=T_i(n)+B+E_i(n)$, where 
$E_i(n)$ is the duration of the $n\,$th back-off time of the
node $X_i$. The non-slotted Aloha principle states that 
$E_i(n)$ are i.i.d. (in $i$ and $n$)
independent of everything else. In what follows we assume that  
$E_i(n)$ are  exponential with mean $1/\epsilon$ 
and will consider the  parameter $\epsilon$ as the 
main parameter to be tuned for non-slotted Aloha (given the
packet emission time $B$). 
More precisely, the lack of
synchronization of the MAC mechanism is reflected in the assumption
that the temporal processes 
$(T_i(n):n\in\bbZ)$ are time-stationary and independent (for different
$i$). Note also that these processes 
are of the {\em renewal} type (i.e., have i.i.d. increments
$T_i(n+1)-T_i(n)$).
For this reason we will call the above case the 
{\em Poisson-renewal model for   non-slotted Aloha}.
The MAC state of the node $X_i$ at (real) time $t\in\bbR$ can be
described by the on-off process 
$e^{renewal}_i(t)=\ind(T_i(n)\le t<T_i(n)+B\; \text{for some\;} n\in\bbZ)$.

\subsection{Fading and External Noise}
\label{ss.Fading}
We need to complete  our network model by some radio channel
conditions. We will consider the following {\em fading
scenario}: channel  conditions 
vary from one transmission to another and between different
emitter-receiver pairs, but remain fixed for any given transmission.
To include this in our model, we assume a further multidimensional mark
$(\bfF_i(n):n\in\bbZ)$ of node $X_i$ where 
$\bfF_i(n)=(F_i^j(n): j)$ with $F_i^j(n)$ denoting the {\em fading}
in the channel from node $X_i$ to the receiver $y_j$ of node $X_j$ 
during the $n\,$th transmission.
We assume that $F_i^j(n)$ are i.i.d. (in $i,j,n$) and independent 
of everything else.
Let us denote by $F$ the generic random variable of the fading.
We always assume that  $0<\E[F]=1/\mu<\infty$.
In the special case of Rayleigh fading, $F$ is exponential (with
parameter $\mu$).  
(see e.g.~\cite[pp.~50 and~501]{TseVis2005}).
We can  also consider non exponential cases, which allow
other types of fading to be analyzed, such as e.g. Rician
or Nakagami scenarios or simply the case without fading (when 
$F\equiv 1/\mu$ is deterministic).

In addition to fading we consider 
vectors $(W_i(n):n\in\bbZ)_i$ of non-negative random
variables, independent in 
$i$ and  of $\widetilde\Phi$, $(\bfF_i(n):i,n)$, 
modeling an external (thermal) noise. More precisely,
$W_i(n)$  models the power of
the external  noise at the receiver $y_i$ at time $n$.
We assume that $W_i(n)$ are identically distributed, and denote
by $\calL_W(s)=\E[e^{-sW}]$ the Laplace transform of the generic 
noise variable $W$. We do not assume any particular
temporal correlation of the noise. In particular, our analysis is
valid for the two extreme cases: of noise $W_i(n)=W_i(0)$ that is
constant in time and noise $W_i(n)$ that is independently re-sampled 
for each time slot $n\in\bbZ$.

The slotted Aloha model described above, when considered in a given
time slot, coincides with the  Poisson Bipolar model with independent
fading considered  in~\cite{JSAC}. 
It allows  an explicit evaluation of the successful transmission
probability and other characteristics such as
the density of successful transmissions, the mean progress, etc. 

An exact analysis of the Poisson-renewal non-slotted Aloha model, 
albeit feasible, does not lead to similarly closed form expressions. To
improve upon this situation,  
in what follows we propose another  model for the non-slotted case.
It allows for the results as explicit  as these 
of~\cite{JSAC}, which are moreover 
very close to these of Poisson-renewal model.

\subsection{Poisson Rain Model for  Non-slotted Aloha}
\label{ss.PoRain}
The main difference with respect to the scenario considered above is
that the nodes $X_i$ and their receives $y_i$ are not fixed in time.
Rather, we consider a time-space Poisson point process
$\Psi=\{(X_i,T_i)\}$ with $X_i\in\bbR^2$ denoting the location of the
emitter which sends a packet during time interval $[T_i,T_i+B)$
(indexing by $i$ is arbitrary and in particular does not mean
successive emissions over time).
We may think of node $X_i$ ``born'' at time $T_n$ transmitting a packet
during time $B$ and ``disappearing'' immediately after.
Thus the MAC state of the node $X_i$ at (real) time $t\in\bbR$ 
is simply  
$e_i(t)=\ind(T_i\le t<T_i+B)$.

We always  assume that $\Psi$ is homogeneous (in time and space)
P.p.p. with intensity $\lambda_s$. This parameter corresponds to the 
{\em space-time frequency of channel access}; i.e, the number of 
transmission initiations  per unit of space and time.
The point $(X_i,T_i)$ of the time-space P.p.p. $\Psi$ are marked by
the receivers $y_i$ in the same manner an described in
Section~\ref{ss.Bipolar}; i.e, given $\Psi$,  $\{X_i-y_i\}$ are i.i.d
random vectors with $|X_i-y_i|=r$.  Moreover, they are marked by 
 $\bfF_i=(F_i^j:j)$, with $F_i^j$ denoting the fading in the channel
from $X_i$ to the $y_j$ 
(meaningful only if $X_i, X_j$ coexist for a certain time).
We assume that  
  $F_j^j(n)$ are i.i.d. (in $i,j$) and of everything else, with 
the same generic random fading  $F$ as in Section~\ref{ss.Fading}.
We will call the above model the {\em Poisson rain model for
  non-slotted Aloha}.  It can be naturally 
motivated by {\em strong mobility of nodes}.

\subsection{Choice of Parameters for the Fair Comparison of the Models}
In order to achieve a fair comparison of the performance of the above 
models, we have to
assume the same offered traffic in all the protocols.
Regarding slotted and renewal non-slotted protocol, we observe that the 
{\em channel-occupation-time-fraction per node} 
(i.e., the average fraction of time each node is authorized to transmit) $\tau$ 
is equal to $p$ in the former and $B/(B+1/\epsilon)$ in the latter one. 
Thus a fair comparison between these two models requires 
\begin{equation}\label{e.p-epsilon-B} 
\tau=p=\frac{B}{B+1/\epsilon}
\end{equation}
and we will consider $p,\tau$ as the main parameters to optimize the
performance of the respective  Aloha models.  
To compare the Poisson rain model the other two models we assume 
the same  {\em space-time density of channel occupation} 
(i.e., the expected total channel-occupation-time by all the nodes,
evaluated per unit of space and time):
\begin{equation}\label{e.ds-p}
\lambda_sB=\lambda \tau\,.
\end{equation} 

\section{Successful Transmission}
\label{s.ST}
\resetcounters
\subsection{Path-loss Model}
Assume that all emitters, when authorized by Aloha, emit packets with
unit signal power and that the receiver $y_i$ 
of node $X_i$ receives a power from the node located at $X_j$
(provided this node is transmitting)
equal to $F_j^i/l(|X_j-y_i|)$,  where \hbox{$|\cdot|$} denotes
the Euclidean distance on the plane and  $l(\cdot)$
is the path loss function.
An important special case consists in taking 
\begin{equation}\label{simpl.att}
l(u)=(Au)^{\beta} \quad \text{for $A>0$ and $\beta>2$.}
\end{equation} 
%
Other possible choices of path-loss function avoiding the
pole at $u=0$ consist in taking e.g. $\max(1,l(u))$,
$l(u+1)$, or  $l(\max(u,u_0))$.

\subsection{SINR Condition}

\subsubsection{Slotted Aloha }
It is natural to assume  that transmitter $X_i$ 
{\em covers} its receiver $y_i$ in time slot $n$ if
\begin{equation}\label{eq:SINR}
\mbox{SINR}_i(n)=\frac{F_i^i(n)/l(|X_i-y_i|)}{W_i(n)+I_i(n)}\ge T\,,
\end{equation}
where  $T$ is some SINR threshold and where
$I_{i}(n)$ is the {\em interference} at receiver $y_i$ at time $n$; 
i.e., {\em the sum of the signal powers 
received by $y_i$ from  all the nodes in 
$\Phi^1(n)=\{X_j\in\Phi: e_j(n)=1\}$ except $X_i$}, namely,
\begin{equation}\label{e.SN-slotted}
I_i(n)=\sum_{X_j\in\Phi^1(n),\, j \ne i }F_j^i(n)/l(|X_j-y_i|)\,.
\end{equation}
When  condition~(\ref{eq:SINR}) is satisfied we say that $X_i$ can be {\em
  successfully  received} by $y_i$ 
or, equivalently,  that $y_i$ {\em is not in  outage} with respect to
$X_i$ in  time slot $n$.

\subsubsection{Non-slotted Aloha}
\label{sss.interference}
When transmissions are not synchronized
(as is the case for non-slotted Aloha) the {\em interference} 
(defined, as previously,  as the sum of the signal powers 
received  by a given receiver from  all the nodes transmitting in the network
except its own emitter) {\em may vary during a
given packet transmission}. Indeed, other transmissions may start or
terminate during this given transmission.
In our Poisson-renewal model of Section~\ref{sss.renewal}
this interference process $I_i(n,t)$
during the $n\,$th transmission to node
$y_i$ can be expressed using~(\ref{e.SN-slotted}) with 
$\Phi^1(n)$ replaced by $\Phi_{ren}^1(t)=\{X_j\in\Phi:
e^{ren}_j(t)=1\}$.  
Similarly, in the Poisson rain model of Section~\ref{ss.PoRain},
the interference process, denoted by $I_i(t)$, 
during the (unique) transmission of node $X_i$ admits the above
representation~(\ref{e.SN-slotted}) with 
$\Phi^1(n)$ replaced by $\Psi^1(t)=\{X_j\in\Psi:
e_j(t)=1\}$, and $F_j^i(n)$ replaced by $F_j^i$.

Below, we propose two different ways of taking into account 
this varying  interference in the SINR condition~(\ref{eq:SINR}):
\begin{itemize}
\item To take  the {\em maximal interference
    value}  during the given transmission
$I_i^{\max}(n)=\max_{t\in [T_i(n),T_i(n)+B]}I_i(n,t)\,$
or $I_i^{\max}=\max_{t\in [T_i,T_i+B]}I_i(t)\,$ for the Poisson-renewal
or the Poisson rain model, respectively; this choice
  corresponds to the situation 
  where bits of information sent within one given packet are not
  repeated/interleaved so that the {\em SINR condition
  needs to  be guaranteed at any time of the  packet transmission}
(for all  
  symbols) for the reception to be successful.     
\item To take  the {\em averaged interference value}
over the whole packet duration
$I_i^{mean}(n)=1/B\int_{T_i(n)}^{T_i(n)+B]}I_i(n,t)\,\md t$
or $I_i^{mean}=\int_{T_i}^{T_i+B}I_i(t)\,\md t$ for the Poisson-renewal
or the rain model, respectively;
 this condition corresponds  to a situation where some coding with
  repetition and interleaving of bits on the whole packet duration is used.
\end{itemize}
More precisely, 
we will say that in {\em non-slotted Aloha with maximal interference
  constraint}
$X_i$ can be successfully  received by $y_i$
(in time slot $n$ in the case of  the Poisson-renewal model), 
if condition~(\ref{eq:SINR}) holds with 
$I_i(n)$ replaced by  $I_i^{\max}(n)$ or $I_i^{\max}$  in the
Poisson-renewal or the Poisson rain 
model, respectively.

Similarly,  we will say that in {\em non-slotted Aloha with average
 interference  constraint}
$X_i$ can be successfully  received by $y_i$
(in time slot $n$ in the case of  the Poisson-renewal model), 
if condition~(\ref{eq:SINR}) holds with 
$I_i(n)$ replaced by  $I_i^{mean}(n)$ or $I_i^{mean}$  in the Poisson-renewal
or Poisson rain model, respectively.

In what follows we will be able to express in closed form expressions 
the  coverage probability for both the Poisson-renewal and the Poisson
rain model when the average interference constraint is considered.
The maximal interference constraint case is studied by simulations
in Section~\ref{ss.mean-max}.

\subsection{Coverage Probability}
In slotted Aloha and the Poisson-renewal model of non-slotted Aloha  
let $\E^0$ denote the expectation with respect to the Palm
probability $\Pro^0$ (cf.~\cite[Sec.~10.2.2]{FnT1}) of the
P.p.p.~$\Phi$. Under this distribution, 
the nodes and their receivers 
are located at $\widetilde\Phi\cup\{(X_0=0,y_0)\}$,  
where $\widetilde\Phi$ is a copy of the original (stationary) marked
P.p.p., and where $y_0$ is independent of $\tilde\Phi$,  
distributed like the other receivers.
Moreover under $\Pro^0$ all other marks of points in $\widetilde\Phi$ and
$X_0$ (MAC status, fading, packet emission renewal processes in the
renewal model) are i.i.d. and have their original distributions.
Under $\Pro^0$, the node $X_0$  at the origin is  called the {\em
  typical node}. (For more details on Palm theory
cf. e.g.~\cite[Sections~1.4, 2.1 and 10.2]{FnT1}.) 
Further  conditioning on the time scale, 
in the case of the slotted Aloha model denote
by $\Pro^{0,e_0=1}\{\,\cdot\,\}=\Pro^0\{\,\cdot\,|\,e_0=1\}$ the
conditional probability of $\Pro^0$ given the  node $X_0$ emits at
time $n=0$. 

In the Poisson-renewal model we denote by $\Pro^{0,T_0(0)=0}=
\Pro^0\{\,\cdot\,|\,T_0(0)=0\}$ the probability $\Pro^0$ given 
the node $X_0=0$ starts transmitting at time $0$.
Formally this means that the renewal
process of transmission times $T_0(n)$ of node $X_0=0$ 
is so called zero-delayed (and we denote by 0 the transmission that starts at
time 0). By
the independence (lack of synchronization)  
other node transmission times are not affected by this conditioning.

Finally let us denote by $p_{slot}$ the {\em probability  $\Pro^{0,e_0(0)=1}$ of
the successful transmission 
of the node $X_0$ at the time 0 given it is selected by the Aloha}; i.e., the  
that the condition~(\ref{eq:SINR}) holds for $i=0$ at time $n=0$.
Similarly, denote by $p^{mean}_{ren}$ the 
{\em probability  $\Pro^{0,T_0(0)=0}$ of
the successful transmission, with the average interference constraint,
of the node $X_0$ started at  time 0 given it is selected by the Aloha}.

In the case of the Poisson rain  model we consider the Palm distribution 
$\Pro^{0,0}$ of the space-time P.p.p. $\Psi$ given a point
$X_0=0,T_0=0$ and denote by $p_{rain}^{mean}$ the {\em probability
  $\Pro^{0,0}$ 
that the transmission from $X_0$ started at time $T_0=0$ is successful
with the average interference constraint}.

Similar notation $p^{\max}_*$ with $*=ren,rain$ will be used for the
probability of successful 
transmission for non-slotted Aloha with the  maximal interference constraint.

\subsubsection{Slotted Aloha}

For the sake of completeness we recall first a result for slotted
Aloha (cf~\cite{JSAC}).
\begin{Prop}\label{prop:outage_M}
Assume the slotted Aloha model of Section~\ref{sss.slotted}
with Rayleigh fading ($F$ exponential with mean $1/\mu$). Then
\begin{eqnarray}
p_{slot}
\!\!\!\!&=\!\!\!\!&\calL_{W}(\mu Tl(r))\label{eq:prlam}\\
&&\!\!\!\!\times\exp \bigg\{-2\pi\lambda p\int_0^\infty
\!\!\!\!\frac{u}{1+l(u)/(Tl(r))}\,\md u\bigg\}.\nonumber
\end{eqnarray}
In particular  if  $W\equiv0$ and that the path-loss
model~(\ref{simpl.att}) is used  then  
\begin{equation}\label{eq:p_r_lambda_1}
p_{slot}=\exp\Bigl\{-\lambda p r^2T^{2/\beta}K(\beta)\Bigr\}\,,
\end{equation}
where
\begin{equation}
\label{eq:kdebeta}
K(\beta)=\frac{2\pi\Gamma(2/\beta)\Gamma(1-2/\beta)}{\beta}=
\frac{2\pi^2}{\beta\sin(2\pi/\beta)}\, .
\end{equation}
\end{Prop}
We remark that the successful transmission probability $p_{slot}$ can also be
evaluated in the case of a general fading distribution $F$;
cf~\cite[Prop.~2.2]{JSAC}. 

\subsubsection{Non-slotted Aloha, Poisson-renewal Model}
Here we present our result for non-slotted Aloha in the  Poisson-renewal model.
Its proof, as well as all other proofs is given in the
Appendix.
\begin{Prop}\label{prop:renewal}
Assume the Poisson-renewal non-slotted Aloha model of Section~\ref{sss.renewal}
with Rayleigh fading ($F$ exponential with mean $1/\mu$). Then 
\begin{eqnarray*}
\lefteqn{p_{ren}^{mean}=\calL_{W}(\mu Tl(r))}\\
&&\times\exp\biggl\{-2\pi\lambda\int_0^\infty u \biggl(1-\frac{1}{1+\epsilon B}\times\Bigl(e^{-\epsilon B}\\
&&+\int_0^B\frac{\epsilon e^{-\epsilon s}}%
{1+\frac{(B-s)Tl(r)}{Bl(u)}}\,\md s\\
&&+\int_0^B\frac{\epsilon}{1+\frac{(B-t)Tl(r)}{Bl(u)}}
\int_0^t\frac{\epsilon e^{-\epsilon s}}%
{1+\frac{(t-s)Tl(r)}{Bl(u)}}\,\md s\md t\Bigr)\biggr)\,\md u\biggr\}\,.
\end{eqnarray*} 
\end{Prop}
As we can see, the above expression for the successful transmission 
in the Poisson-renewal model, albeit numerically tractable, is not
very explicit.  
In the following section we show that the Poisson rain model
leads to much more tractable results.

\subsection{Non-slotted Aloha --- Poisson Rain Model}
Here we present our main result for this model.
\begin{Prop}\label{p.PoRain}
Assume the Poisson Rain model for  non-slotted Aloha  of
Section~\ref{ss.PoRain} 
with Rayleigh fading ($F$ exponential with mean $1/\mu$).
Then
\begin{eqnarray}
p_{rain}^{mean}
\!\!\!\!&=\!\!\!\!&\calL_{W}(\mu Tl(r))\label{eq:PoRain}\\
&&\hspace{-3em}
\times\exp \bigg\{-4\pi\lambda_sB\int_0^\infty
\!\!\!\! u\Bigl(1-\frac{l(u)}{l(r)}\log\Bigl(
1+\frac{l(r)T}{l(u)}\Bigr)\,\md u\biggr\}\,.
\nonumber
\end{eqnarray}
In particular  if  $W\equiv0$ and that the path-loss
model~(\ref{simpl.att}) is used  then  
\begin{equation}\label{eq:p_rain}
p_{rain}^{mean}=\exp(-\lambda_sB r^2T^{2/\beta}K'(\beta))\,,
\end{equation}
where
\begin{equation}
\label{eq:k_rain}
K'(\beta)=\frac{4\pi}{\beta}
\int_0^\infty u^{2/\beta-1}(1-u\log(1+u^{-1}))\,\md u\,.
\end{equation}
\end{Prop}
The successful transmission probability
$p_{rain}^{mean}$ can also be effectively 
evaluated in the case of a general fading distribution~$F$; see Appendix.

\rem 
We consider the Poisson-rain model to be a simplified model for non-slotted
Aloha. In particular it has only one parameter $\lambda_s$ (time-space
density of transmission initiations) that does not allow us
to distinguish between the spatial density  $\lambda$ of
nodes  and the channel-occupation-time-fraction per node.
However, in Section~\ref{ss.Validation} we validate this model 
by comparing the probability of successful
transmission $p_{rain}^{mean}$ to $p_{ren}^{mean}$  
under equality~(\ref{e.ds-p}) with $\tau=B/(B+1/\epsilon)$. 
We will see a very good  matching.
Given this observation we can  use~(\ref{eq:PoRain})
with $\lambda_sB=\lambda\tau$ to express  the 
basic performance metric  of the non-slotted
Aloha (probability of successful transmission) in terms of all the 
parameters of the (real) non-slotted system. 
In the case of $W=0$ and the path-loss function~(\ref{simpl.att})
this relation has the following simple form  
\begin{equation}\label{e.p_c_non-slotted}
p_{ns}=\exp\Bigl\{-\frac{\lambda}{1+1/(\epsilon B)}
r^2T^{2/\beta}K'(\beta)\Bigr\}\,. 
\end{equation}

\subsection{Slotted Versus Non-slotted Aloha --- First Comparison}
Note that the expression in~(\ref{eq:p_rain}) has exactly the same form as
that for the slotted Aloha in~(\ref{eq:prlam}),
provided~(\ref{e.ds-p}) holds (i.e., when the both schemes exhibit the same
space-time density of channel occupation, with the only
difference being in the path-loss dependent constant $K'(\beta)$. 
This observation allows for an explicit comparison of several 
performance metrics of the slotted and non-slotted Aloha.
The simplest one consists in comparing the blocking probabilities 
$p_{slot}$ to $p_{ns}$ given the same tuning of both systems.
\begin{Res}
Assume the  same density of nodes $\lambda$,
transmission distance $r$,  and 
the same channel-occupation-time-fraction per node  $\tau$~(\ref{e.ds-p}).
In the case of Rayleigh fading the non-slotted Aloha offers 
$$\frac{p_{ns}}{p_{slot}}\times 100\%=
e^{-(K'(\beta)-K(\beta))\lambda r^2T^{2/\beta}\tau}\times 100\%$$
of the good-put (frequency of the successful transmissions)
per node  of the slotted Aloha.
\end{Res}

\begin{figure}[!t]
\centering\includegraphics[width=0.9\linewidth]{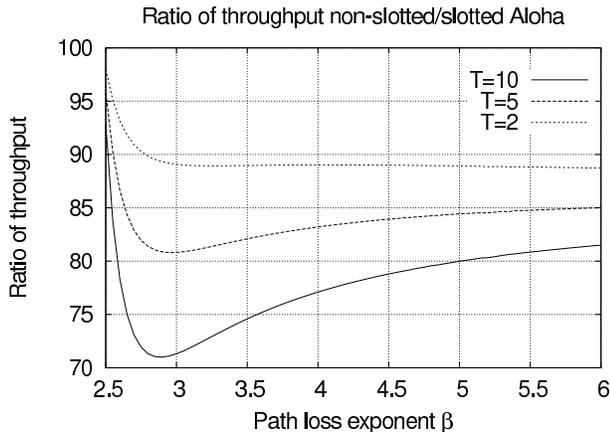}
\caption{The ratio (in \%) of the good-put offered by the non-slotted
  Aloha with respect to the slotted one, as a function of the path loss
  exponent $\beta$, for various choices of the SINR threshold $T$;
other parameters are $p=\frac{\epsilon}{1+\epsilon}=0.05$, $\lambda = 0.001$, $r=\sqrt{1000}$.}
\label{fig.cpdebit_bvar}
\end{figure}

In Figure~\ref{fig.cpdebit_bvar}, the ratio 
$\frac{p_{ns}}{p_{slot}}\times 100\%$ 
is shown for different values of the path-loss exponent~$\beta$ 
and SINR threshold~$T$. For other parameters we take
$p=\frac{\epsilon}{1+\epsilon}=0.05$, $\lambda = 0.001$, $r=\sqrt{1000}$. 
For $T=10$ and $\beta$ close to $2.5$ slotted and non-slotted Aloha
have similar performances.  
For higher values of $\beta$ non-slotted Aloha offers a good-put 
ranging from $70\%$ to $80\%$ of this of slotted Aloha.
Note that the above comparison concerns performance of the non-optimized (in $p$ and $\tau$) schemes. We compare both models under
their respective optimal tuning in what follows.

\section{Optimal Tuning of Non-slotted Aloha}
\label{s.tuning}
\resetcounters
In what follows we are interested the following  MANET 
performance metrics introduced in~\cite{JSAC}
($p_c$  denotes $p_{slot}$, or $p_{ns}$ in the slotted or non-slotted
 Aloha case, respectively):
\begin{itemize}
\item {\em (space-time) density of successful transmissions}
$d_{suc}=\lambda\tau p_c$,
\item {\em mean progress}
$prog=r\, p_c$.
\end{itemize}

\subsection{Optimal MAC for the Density of Successful
  Transmissions}
Assume $\lambda, r$ to be fixed.
A good tuning of the non-slotted Aloha renewal parameter $\epsilon$
(or equivalently of  $\tau$,  given $B$) 
 should find a compromise between the average number of
concurrent transmissions per unit area and the probability that a
given authorized transmission will be successful.
To find such a compromise,  one can e.g. 
maximize the time-space  frequency of  successful transmissions $d_{suc}$.
The following result follows immediately
from~(\ref{e.p_c_non-slotted}).
\begin{Res}\label{r.opt_d_succ}
Assume no noise $W=0$,  Rayleigh fading and
 path-loss~(\ref{simpl.att}). Given $r$, 
the  maximum value of the  
density of successful transmissions $d_{suc}=1/(eK'(\beta)r^2T^{2/\beta})$ in the non-slotted Aloha 
is attained for the space-time density of channel access
$\lambda\tau=1/(K'(\beta)r^2T^{2/\beta})$.
Moreover, given the spatial density of nodes
$\lambda$ the optimal mean channel-access-time-fraction $\tau$ per
node 
$\tau_{\max}$ for  $d_{suc}$
is equal to 
$$\tau_{\max}=\frac1{\lambda K'(\beta)r^2T^{2/\beta}}\,.$$
if $\lambda>1/(K'(\beta)r^2T^{2/\beta})$
and $\infty$ (interpreted as no back-off; i.e., immediate retransmission)
otherwise.
\end{Res}

\rem
Recall from~\cite{JSAC} that  similar optimal value of $d_{suc}=1/(e
K(\beta)r^2T^{2/\beta})$ 
for slotted Aloha is attained for 
$\lambda p_{\max}=1/(K(\beta)r^2T^{2/\beta})$.
Since $K'(\beta)>K(\beta)$ we have $p_{\max}>\tau_{\max}$, which means
that optimally tuned  non-slotted  Aloha occupies less channel
than optimally tuned slotted Aloha. However, both schemes
exhibit the same  energy efficiency. Indeed 
if one assumes that each  transmission requires a unit 
energy, the number of successful transmissions per unit of 
energy spent is $1/e$. Remark that the latter comparison does not take into
account the energy spent to maintain synchronization in the slotted scheme.

\begin{figure}[t]
\centering\includegraphics[width=0.90\linewidth]{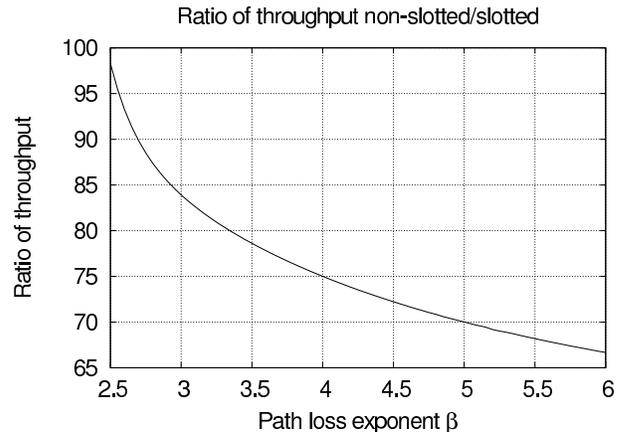}
\caption{The ratio (in \%) of the good-put offered by the non-slotted
  Aloha with respect to the slotted one, when both are optimized so as
  to  to maximize the density of successful transmissions as a function 
of the path loss exponent $\beta$. This ratio does
not depend on any other parameter.} 
\label{fig.cpdebitop_bvar}
\end{figure}

The following result compares the optimal density of transmission in 
slotted and non-slotted Aloha.
\begin{Res}
Under the assumptions of Result~\ref{r.opt_d_succ},
 for a given density of nodes $\lambda$
non-slotted Aloha with the optimal tuning $\tau_{max}$
offers 
$$\frac{\tau_{\max}p_{ns}(\tau_{\max})}{p_{\max}p_{slot}(p_{\max})}\times 100\%=
\frac{K(\beta)}{K'(\beta)}\times 100\%$$     
of the good-put 
of the optimally tuned slotted Aloha.
\end{Res}
In Figure~\ref{fig.cpdebitop_bvar} we present this good-put ratio
for the optimized systems in function of $\beta$
(note that it does not depend on other parameters like $\lambda,r,T$).   
We observe that for small values of path-loss exponent $\beta$ (close
to $2$ the performances of slotted and non-slotted Aloha are similar but for
large values of $\beta$ non-slotted Aloha performs significantly worse 
than the slotted one.
In fact, more extensive numerical  computations (not presented here)
allow us to conjecture that 
$\lim_{\beta\to 2+}K(\beta)/K'(\beta)=1$ and $\lim_{\beta\to
    \infty}K(\beta)/K'(\beta)=0.5$.
The second part of this conjecture means that the
good-put ratio for the optimized systems only asymptotically,
when $\beta\to\infty$,  goes to $0.5$  --- the value predicted 
by the widely used simplified model  with the simplified collision model 
(see \cite[Section~4.2]{Bertsekas01}).
However, e.g. for $\beta=4$ 
this ratio is still~$75\%$ and even for $\beta=6$ the ratio still
remains significantly larger than $50\%$.   

When trying to explain the above asymptotic value of~50\%,
one may argue that in the presence of a very strong path-loss the only
significant interferers are those, closer to the given 
receiver than its own emitter, and that their
impact is the same as if they were all located in an immediate
vicinity to the receiver. This makes the channel between a given
emitter and its receiver compatible with the classical, ``geometryless'' model.

\subsection{Optimal Transmission Distance Given MAC}
We assume now that the density of nodes $\lambda$ as well as
some tuning of MAC ($\tau$) is given. 
We are interested in finding the transmission distance $r$ 
that maximizes the mean  progress~$prog$ in the network.
The following result follows immediately from~(\ref{e.p_c_non-slotted}).

\begin{Res}\label{r.opt_prog}
Assume no noise $W=0$,  Rayleigh fading and
 path-loss~(\ref{simpl.att}). Given $\lambda$ and $\tau$, 
the  maximum mean progress $prog$
in the non-slotted Aloha 
is attained for the transmission distance 
\begin{equation}\label{e.r_max}
r_{\max}=\frac{1}{\sqrt{2K'(\beta)T^{2/\beta}\lambda\tau}}
\end{equation}
\end{Res} 
Recall from~\cite{JSAC} that the similar optimal tuning of $r$ in
slotted Aloha is equal to~(\ref{e.r_max}) with $K'(\beta)$ replaced
by~$K(\beta)$. It is thus larger than this for non-slotted Aloha.
Here is the comparison of the mean progress in both systems.
\begin{Res}
Under the assumptions of Result~\ref{r.opt_prog}
the non-slotted Aloha with the optimal transmission distance 
offers a mean progress
$$\frac{prog_{ns}(r_{\max})}{prog_{slot}(r_{\max})}\times 100\%=
\frac{\sqrt{K(\beta)}}{\sqrt{K'(\beta)}}
\times 100\%$$
of the mean progress in the slotted Aloha 
with the optimal transmission distance and the
same tuning of MAC $\tau=p$.
\end{Res}

The curve corresponding to this result is very similar to the one 
presented in Figure~\ref{fig.cpdebitop_bvar}. In particular for $\beta=4$, the 
ratio of the mean progress is close to $87\%$.


\section{Further Numerical Results}
\resetcounters
\subsection{Validation of the Poisson Rain Model}
\label{ss.Validation}

\begin{figure}[t]
\centering\includegraphics[width=0.90\linewidth]{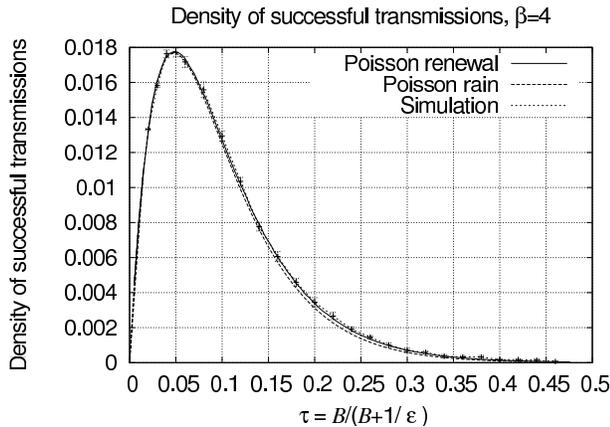}
\caption{Density of successful transmissions versus 
$\tau=\frac{B}{B+1/\epsilon}$. Comparison of the Poisson-renewal and the Poisson rain model to simulation results.}
\label{fig.ns_aloha_modsim}
\end{figure}

In this section we validate our Poisson rain model by comparing
the successful transmission probability  $p^{mean}_{rain}$
to the same characteristic $p_{ren}^{mean}$ evaluated numerically 
with the formula of Proposition~\ref{prop:renewal} for 
our Poisson-renewal model of non-slotted Aloha. 
We use the same numerical assumptions as previously:
$\lambda = 0.001$, $r=\sqrt{1000}$, 
$T=10$ and $\beta=4$. 
We also compare the results of these  two models with 
simulations carried out in a square of 1000 m $\times$ 1000 m with 
the same numerical assumptions; this is shown 
in Figure~\ref{fig.ns_aloha_modsim}. We observe an excellent 
matching of the two models with simulations, the differences 
being almost imperceptible. \footnote{The error bars in all
  simulation results correspond to a
confidence interval of $95\%$.}
We perform the same comparison between the two models and the simulations 
for $\beta=5$ and $\beta=3$. For $\beta=5$ the matching of the 
two models and the simulation is perfect. For $\beta=3$ the two 
models provide the same results whereas the simulations give a larger 
density of throughput. This can be explained by the fact the
analytical models regard infinite-plane models, while in simulations 
the network area  is  finite, and the border effects have stronger
impact  for small values of $\beta$.

\subsection{Mean Versus Maximum Interference Constraint in SINR}
\label{ss.mean-max}
In this section we show the impact of the 
assumption on maximum interference
constraint in the SINR
on the probability of a successful transmission.

\begin{figure}[t]
\centering\includegraphics[width=0.90\linewidth]{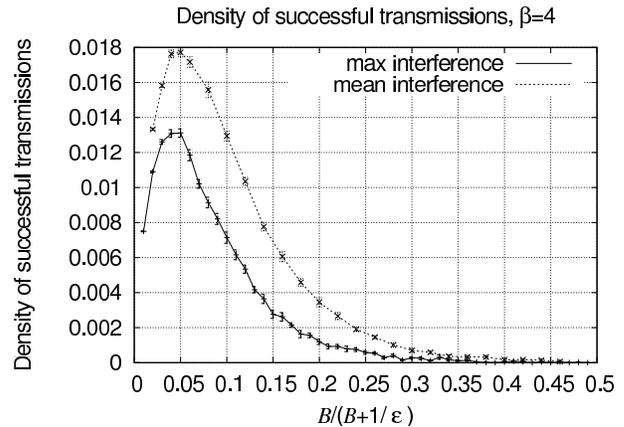}
\caption{Density of successful transmissions versus
  $\tau=\frac{B}{B+1/\epsilon}$ for mean and maximal interference constraint; simulation
  results. }
\label{fig.ns_aloha_mean_max_b4}
\end{figure}

In Figure~\ref{fig.ns_aloha_mean_max_b4} we compare $p^{max}_{ren}$ to 
$p_{ren}^{mean}$  for   $\lambda = 0.001$, $r=\sqrt{1000}$, 
$T=10$ and $\beta=4$.
The loss in performance when the SINR is computed with the maximum 
interference can be large and may be up to $45\%$. But when the 
throughput is optimized in $\epsilon$, the loss in performance is only  
$26\%$. We observe that the throughput is optimized in both 
cases for the same value of $B\epsilon \simeq 0.045$, this value 
is also optimal for the Poison rain model.
The density of successful transmissions for 
non-slotted Aloha when the SINR is not averaged is  $55\%$ 
that of slotted Aloha. In this case, the comparison is  close to 
those of the 'standard' model of slotted/non-slotted Aloha on a 
wired network. 

\begin{figure}[!h]
\centering\includegraphics[width=0.90\linewidth]{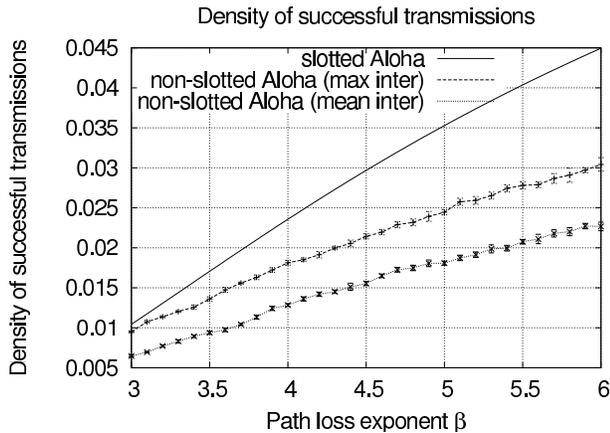}
\caption{Density of successful transmissions versus path loss exponent
 $\beta$. Slotted Aloha and non-slotted Aloha (mean and maximal interference constraint) are tuned to maximize the density of successful transmissions. }
\label{fig.ns_aloha_mean_max_bvar}
\end{figure}

In Figure~\ref{fig.ns_aloha_mean_max_bvar} we compare the density of 
successful transmissions for slotted Aloha and non-slotted Aloha when 
the maximum or average SINR is considered. For slotted Aloha we use 
the analytical model and optimize the  density of throughput in $p$. 
For non-slotted Aloha we use simulation results and the Poison rain 
model to optimize the schemes in $\frac{B}{1+1/\epsilon}=\tau$. We observe 
that for $\beta \leq 4$  non-slotted Aloha with the averaged SINR 
provides $50\%$ more throughput than with the maximum SINR.  
For $\beta \geq 5$ non-slotted Aloha with the averaged SINR 
provides only around $35\%$ more throughput than with the maximum SINR.
When we compare slotted Aloha with non-slotted Aloha with maximum 
SINR, we find that slotted Aloha offers $66\%$ more throughput for 
$\beta=3$ and $100\%$ for $\beta=6$.

\section{Conclusion}
\resetcounters
\label{s.conclusion} 
We have developed two stochastic  models to analyze non-slotted 
Aloha in SINR based scenarios.
If we consider Rayleigh fading, a power-law signal-power 
decay and the interference 
to be averaged over the duration of the transmission slot, 
our two models 
lead to closed formulas for the probability of capture and the 
density of throughput. 
The formula of the Poisson rain model 
can be very simply used to provide 
straightforward results whereas the Poison-renewal model 
requires more computational effort. 
The two models also provide very close 
results, which are confirmed by simulations. 
The analysis can also be extended for a general 
fading $F$ using the Plancherel-Parseval theorem.

Our two models for non-slotted Aloha allow extensive performance 
comparisons with slotted Aloha. We compare non-slotted 
and slotted Aloha both for a given value of transmission attempts 
as well as when the two models are optimized. Slotted Aloha does indeed 
offer better performances than non-slotted Aloha but for realistic 
path-loss assumptions the ratio
is far smaller than the  ``well-known''  factor~2 obtained 
for Aloha in the wired model,

Using the simulation results we have studied non-slotted Aloha when
the maximum value  
rather than the average value of
the interference is considered in the SINR constraint. 
This change introduces a significant loss in the density of throughput 
but this loss is greatly reduced when the density of successful
transmissions is optimized. 


\appendix
\label{s.appendix} 
\renewcommand{\theequation}{A.\arabic{equation}}
\section{Appendix}
In this section we prove Propositions~\ref{p.PoRain}
and~\ref{prop:renewal},
and show some extensions of these results.

\subsection{General Approach}
We begin with the simple observation that the successful transmission
probability $p_c$ can be expressed in all cases considered in this paper
(slotted Aloha, Poisson-renewal and Poisson rain, both with maximum or
average interference constraint) in therms of the following {\em
  independent} random variables 
$$p_c=\Pro\{\,F \ge T l(r)(I+W) \,\}\,$$
where $F$ and $W$ are the generic variables  representing,
respectively, fading and external
noise, and where $I$ is the appropriate interference (maximum or averaged
during the reception of the given packet in the non-slotted case;
cf. Section~\ref{sss.interference}). 
Thus, in the case of Rayleigh fading, we have.
\begin{Fact}\label{f.Reyleigh}
Assume exponential $F$ with mean $1/\mu$ (Rayleigh fading). Then 
$$p_c=\E[e^{-\mu T l(r)(I+W)}]=\calL_W(\mu T l(r))
\calL_I(\mu T l(r))\,,$$
where $\calL_I(\xi)$ is the Laplace transform of the 
interference in the respective model.
\end{Fact}
The above formula was used for the first time in~\cite{allerton} in the
case of slotted Aloha.
The following result proved in~\cite{JSAC} (for slotted
Aloha)
allows the analysis to be extended to the case of a general fading $F$.
It follows from the Plancherel-Parseval theorem (see
e.g.~\cite[Th.~C3.3, p.157]{Bremaud2002}).
\begin{Fact}\label{f.General}
Assume that 
\begin{itemize}
\item $F$ has a finite first moment and admits a square integrable density;
\item Either $I$ or $W$ admit a density which is square
integrable~\footnote{%
The square integrability of the density of a given random variable (in
particular of~$I$) 
is equivalent to the square integrability of its Fourier transform 
(in particular to the integrability  of $|\calL_{I}(is)|^2$ in the
case of~$I$); see~\cite[p.510]{FellerII}.}
\end{itemize}
Then the probability of a
successful transmission is equal to 
\begin{eqnarray*}
\lefteqn{p_c}\\
 &&\hspace{-2em} =\!\!
\int_{-\infty}^{\infty}
{\calL}_{I}\!
\left(2i\pi l(r)Ts\right)
{\calL}_W\!
\left(2i\pi l(r)Ts\right)
\frac{{\cal L}_F(-2i\pi s)-1}{2 i\pi s} \md s\,,\nonumber
\end{eqnarray*}
where $\calF(\xi)=\E[e^{-\xi F}]$ is the Laplace transform of~$F$.
\end{Fact}
The results of Propositions~\ref{p.PoRain}
and~\ref{prop:renewal} follow now from Fact~\ref{f.Reyleigh}
and the particular form of the Laplace transform of the corresponding
{\em averaged interference} $I=I^{mean}$ of the typical user transmission 
in the Poisson-renewal and Poisson rain model. 
We are unable to give an analytical expression for the
successful transmission in either of the non-slotted Aloha models
under the {\em maximal interference constraint} due to the fact that
we are not aware of any explicit representation of the Laplace
transform of the maximal interference in the considered models.

In what follows we develop expressions  for the Laplace
transforms of $I^{mean}$. 
We use the following result giving an explicit
formula for the Laplace transform of the  {\em generic shot-noise}
$J=\sum_{Y_i\in\Pi}f(G_i,Y_i)$ generated by some homogeneous Poisson
p.p. with intensity $\alpha$, the  {\em response function}
$f(\cdot,\cdot)$ and i.i.d. (possibly multi-dimensional) marks $G_i$
distributed as a generic r.v. $G$.
 It can be derived
from the formula for the Laplace functional of the Poisson p.p.
(see e.g.~\cite{daley}).
\begin{Fact}\label{f.I_SN}
Consider the shot-noise random variable defined above.
Then 
\begin{equation}
\calL_{J}(s)=\E[e^{-sJ}] \label{e:JLT}
=\exp\Bigl\{-\alpha \int(1-\E[\exp\{-s f(G,y)\}])\,\md y\Bigr\}\,, 
\end{equation}
where the integral is evaluated over the whole state space of on which
P.p.p. $\Pi$ lives and the expectation $\E$ in the exponent is taken
with respect to the distribution of the generic mark~$G$.
\end{Fact}

\subsection{Interference in the Poisson Rain Model}
We begin with the simpler --- Poisson-rain  case.
Recall that, in this case, we have 


$$
I^{mean}=1/B\int_0^B\sum_{X_j\in\Psi^1(t)\, j \ne 0}F_j^0/l(|X_j-y_i|)\,\md t
$$

considered under
the Palm probability $\Pro^{0,0}$ of the space-time P.p.p. $\Psi$ given a point
$X_0=0,T_0=0$.
By the stationarity of $\Psi\setminus\{X_0\}$ under   $\Pro^{0,0}$
we can replace $y_i$ by $0$ in the above formula. Moreover using the
representation $\Psi^1(t)=\{X_j:1=e_j(t)=\ind(T_j\le t<T_j+B)\}$
for nodes that emit at time $t$, changing
the order of integration and summation we obtain the following
representation  for (the distribution of) $I^{mean}$ 
\begin{equation}\label{e.I_rain}
I^{mean}=_{distr.}
\sum_{(X_j,T_j)\in\Psi}F_jh(T_j)/l(|X_j|)\,,
\end{equation}
where $\Psi$ is the stationary space-time P.p.p. of the Poisson rain
model, $F_j$ are i.i.d. copies of the fading variable $F$ 
and 
\begin{equation}\label{e.h}
h(s)=\int_0^B
\hspace{-0.5em}\frac{\ind(s\le t<s+B)}B\,\md t=
\frac{(B-|s|)^+}{B}\,,
\end{equation}
where $a^+=\max(0,a)$.
Note that the random variable on the right-hand-side
of~(\ref{e.I_rain}) is an example of the shot-noise random variable 
$J$, with respect to the P.p.p. $\Psi$ on $\bbR^2\times\bbR$
 with the response function 
$f(F,(x,t))=F\,h(t)/l(|x|)$. Using Fact~\ref{f.I_SN}
we can obtain an explicit formula for the Laplace transform of
$I^{mean}$ in this case. For the sake of simplicity, below we only give it 
below for the case of
Rayleigh fading.
\begin{Fact}\label{f.SN-Po}
The Laplace transform of the averaged interference during the packet
reception of the typical packet in  the Poisson rain model with the
Rayleigh fading (exponential $F$ with mean $1/\mu$) is equal to 
$$\calL_{I^{mean}}(\xi)=
\exp\Bigl\{
-4\pi \lambda_sB\!\!\int_0^\infty \hspace{-1.3em}u
\Bigl(1-\frac{\mu l(u)}{\xi}\log(1+\frac{\xi }{\mu l(u)})\Bigr)\,\md u\Bigr\}\,.
$$
\end{Fact}

\subsection{Interference in the Poisson-Renewal Model}

As for the above demonstration, 
we consider the transmission at $X_0=0$ starting 
at time $T_0=0$ under the $\Pro^{0,T_0(0)=0}$. Each node  $X_j$ 
can send at most two packets interfering with the given transmission.
To identify them let us denote by $n_j^*$ the unique integer such that
$T_j(n_j^*) \leq 0< T_j(n_j^*+1)$.  To simplify the notation denote
$R_j= T_j(n_j^*)$ and $S_j=T_j(n_j^*+1)$. These are the arrival times
of the two (potentially) interfering packets transmitted by $X_j$.
We have for the distribution of the mean interference $I^{mean}$ 
at node $X_0=0$:  
\begin{equation}\label{e.I_renewal}
I^{mean}=_{distr.} \sum_{X_j\in\Phi}F_jh(R_j)/l(|X_j|)+ F_{j}'h(S_j)/l(|X_j|)  \,,
\end{equation}
where $(F_j,F_j':j)$ are independent copies of the fading variable
$F$, independent of $(R_j,S_j)$.
Using Fact~\ref{f.I_SN} we obtain  the Laplace transform of the above 
shot-noise variable:  
\begin{eqnarray*}
\calL_{I^{mean}}(\xi)& = &\exp\Bigl\{ -\lambda\int_{\mathbb{R}^2} 
\Bigl(1-\E^{0,T_0(0)=0}\Bigl[\frac{\mu}{\mu+\xi \frac{h(R)}{l(|x|)}}   \\
&& \times \frac{\mu}{\mu+\xi \frac{h(S)}{l(|x|)}}\Bigr]\Bigr)\md
x\Bigr\}, 
    \end{eqnarray*}
where the expectation is with respect to $(R,S)$ --- 
a generic copy for $(R_j,S_j)$   
(the expectation with respect to  the exponentially 
distributed fading variables $F,F'$ 
has already been taken into account in the formula).

According to renewal theory (see
e.g.~\cite[eq.~1.4.3]{BaccelliBremaud2003}),  
the joint distribution of $(R,S)$ is given by $\Pro\{\,-R+S\le B+a\,\}
=\int_0^a\frac{(B+s)\epsilon}{B+1/\epsilon} e^{-\epsilon s}\,\md s$
for $a\ge0$.
From this formula we derive first the marginal law of $R$
that is with probability 
$\frac{\epsilon B}{1+ \epsilon B}$ uniformly distributed on $[-B,0]$
and with probability $\frac{1}{1+ \epsilon B}$ equal to $-(B+e_{\epsilon})$ 
where $e_{\epsilon}$ is an exponentially 
distributed random variable of rate $\epsilon$. 
Next, the  conditional distribution of $S$ given $R$ can be derived:
if $R \geq -B $ then $S = B+R+e_{\epsilon}$ and 
$R = e_{\epsilon}$ otherwise. Using these distributions, the
required expectation can easily computed:  
\begin{eqnarray*}
\lefteqn{ \calL_{I^{mean}}(\mu Tl(r) ) = \exp\Bigl\{-\lambda \int_{x\in\mathbb{R}^2}\Bigl(1-\frac{\epsilon B}{1+ \epsilon B} } \\
&&
\times \frac{1}{B} \int_0^B \frac{1}{ 1+\frac{(B-t)Tl(r)}{Bl(|x|)}} \int_0^{\infty} \frac{\epsilon e^{-\epsilon s}}{ 1+\frac{(t-s)^+Tl(r)}{Bl(|x|)}} \md s \md t \\
&& 
- \frac{1}{1+ \epsilon B} \int_0^{\infty} \frac{\epsilon e^{-\epsilon s}}{ 1+\frac{(B-s)^+Tl(r)}{Bl(|x|)}} \md s    \Bigr)\md x  \Bigr\}\,.     
\end{eqnarray*}
After some simplifications and a change in polar variables 
this gives the result announced in Proposition~\ref{prop:renewal}.

\singlespacing
\bibliographystyle{plainnat}
{\footnotesize 
\bibliography{aloha}
}

\end{document}